# Josephson junctions and superconducting quantum interference devices made by local oxidation of niobium ultrathin films

V. Bouchiat, M. Faucher
GPEC, UMR CNRS 6631, Université de la Méditerranée, Case 901,
13288 Marseille, France.

C. Thirion, W. Wernsdorfer
Laboratoire Louis Néel, UPR CNRS 5051, BP 166, 38042 Grenoble, France.

T. Fournier, B. Pannetier
CRTBT, UPR CNRS 5001, BP 166, 38042 Grenoble, France.

**ABSTRACT**

We present a method for fabricating Josephson junctions and superconducting quantum interference devices (SQUIDs) which is based on the local anodization of niobium strip lines 3 to 6.5 nm-thick under the voltage-biased tip of an Atomic Force Microscope. Microbridge junctions and SQUID loops are obtained either by partial or total oxidation of the niobium layer. Two types of weak link geometries are fabricated : lateral constriction (Dayem bridges) and variable thickness bridges. SQUIDs based on both geometries show a modulation of the maximum Josephson current with a magnetic flux periodic with respect to the superconducting flux quantum *h/2e*. They persist up to 4K. The modulation shape and depth for SQUIDs based on variable thickness bridges indicate that the weak link size becomes comparable to the superconducting film coherence length ? which is of the order of 10nm.

**Corresponding author** :   Vincent  Bouchiat
                            Tel : +33 4 91 82 92 16
                            Fax : +33 4 91 82 91 76
                            e-mail : bouchiat@gpec.univ-mrs.fr

Nanolithography using scanning probe microscopes (SPM) offers powerful methods [1] for patterning surfaces with a resolution beyond the range of conventional lithographies based on resist exposures. During the last decade, these techniques have brought some important features to device fabrication, such as easy alignment and *in-situ* control of the device electrical characteristics during its fabrication [2]. Furthermore, local probe techniques based on near field interactions, show a greatly reduced proximity effect [3] which limits resolution in e-beam lithography. On the one hand, lithography using UHV-STMs lead to ultimate resolution (i.e. at the atomic scale) [4-5], but the building of structures with a permanent and stable electrical connection [6] has not been completely achieved. On the other hand, non-UHV SPM lithography techniques are mainly based on the atomic force microscope (AFM). These latter techniques, based either on tip indentation [7-9] or on a voltage biased tip [10-19], retain nanometer scale resolution and show a better versatility. Among these resist-less AFM lithographies, local anodisation of the surface of a semiconductor [11,12] or of non-noble metals [13-15] by the biased tip of an AFM is a versatile method for making nanoscale quantum devices. Quantum point contacts [11,13], nanowires [12], single electron devices [14], superconducting devices [16] as well as other nanoscale devices involving nanotubes [17] or clusters [18] have been obtained.

In order to fabricate the structure in a single step, the film thickness must be less than the typical depth of oxidized metal (i.e 10 nm), thus allowing a direct writing of fully insulating regions. Such a process can be well controlled and is sufficiently reproducible to control the oxide linewidth to values defining either a complete electrical separation or, for a single line drawn at high speed and low voltage, a metal/insulator/metal tunnel barrier with low transparency [15]. The intrinsic ultra-small capacitance of such tunnel junctions has been exploited to produce single electron devices operating at room temperature [14].

In this letter, we present an application of this anodisation technique for fabricating superconducting nanostructures using high quality ultra-thin niobium films. As a first demonstration of potential applications for mesoscopic superconductivity, we have made and tested at low temperature a series of superconducting quantum interference devices (DC-SQUIDs) based on the microbridge technology with various geometries.

A single crystal sapphire wafer was chosen as the substrate for ultra-thin film growth. It has a $\bar{1}102$ orientation with an off-axis miscut as low as possible (about $10^{-3}$ rad). After a thermal treatment at 1100°C for 1h in air [19], the sapphire surface is reconstructed such that 0.3-0.8µm wide, atomically flat terraces separated by 0.3 nm high steps are observed (visible in fig.1b). A niobium layer of thickness either 3 or 6.5 nm is then epitaxialy grown with using an electron gun evaporator in UHV conditions. The sapphire substrate was in-situ cleaned by Ar ion milling and heated at 550°C during Nb deposition [20]. In order to prevent rapid aging of the film, a 2-nm-thick silicon layer is deposited on top at room temperature in the same vacuum [21]. As expected, the films show superconducting properties slightly depressed with respect to the bulk [20]: critical temperatures of the bare films are respectively 5 K and 6.6 K, while their residual resistivity ratios (T=300K /T=4.2K) are 1.5 and 2.2. [22].

Before proceeding to the AFM lithography, a pre-fabrication step is performed in order to define the electrical connections. The film is patterned using standard UV lithography techniques and is dry etched in an $SF_6$ plasma to define 1 to 3 µm wide strip lines. Their electrical resistance is of the order of 60? per square for the 6.5nm-thick Nb layer. The wafer is then diced and AFM lithography is performed using the contact mode with a commercial PtIr-covered tip.
A negative voltage ranging between 4 and 14 Volts is applied on the tip with respect to the grounded film. This voltage depends on both the desired oxide depth and the tip quality. We have found that the silicon top layer does not affect the oxidation of the underneath Nb layer.

Large insulating areas are obtained by scanning the biased tip in lines laterally separated by 10 to 40 nm. Since tunneling barriers obtained by a single oxide line diffusing through the whole layer thickness are still too thick for enabling a measureable Josephson tunneling current [14], our Josephson junctions are based on superconducting weak-links [23]. The idea is simply to pattern two local constrictions (microbridges) on the strip line separated by the insulating loop center. Figure 1 shows AFM images of two types of SQUIDs chosen from the ~40 devices fabricated. Microbridges can be clearly seen on both sides of the strip line: oxidized areas which are thicker than the bare niobium film appear bright in AFM micrographs. Two types of microbridge geometries have been successfully tested. First we have fabricated SQUIDs with so-called "Dayem" bridges [24]. These bridges consist of superconducting lateral constrictions of widths between 30 and 100nm and length 200-1000 nm [Fig. 1(a)]. Their resistance in the normal state is estimated to be of several hundred ohms.

The figure 2 presents the electrical characterization at low temperature of the sample imaged in Fig.1a. Switching currents $I_s$ (see arrow in Fig. 2) are defined as maximum DC-Josephson currents measured before reaching the non-zero-voltage dissipative regime. Non-linearities at finite voltage come from heat induced transitions : their positions depends on the current sweeping frequency.

In the AFM-made Dayem-like SQUIDs presented in Fig. 1a, the modulation of $I_s$ with magnetic field exhibits a sharp symetric saw-tooth shape with modulation depth ?$I_s$/$I_s$ of 12-18%, where ?$I_s$ is the peak-to-peak modulation depth. Period was reproducibly found to be around 2mT (Fig. 2, top inset), a value in good agreement with the predicted period of ?$_?$/S, where S is the loop area (~1µm²) and ?$_?$ the superconducting flux quantum $h/2e$. The linear dependence of $I_s$(? ), as well as the reduced modulation depth could suggest that the loop screening currents are strong enough to partially wash out the interference pattern. However the calculated loop inductance L of our device (2pH) leads to a reduced screening factor $LI_s$?$_0$=0.003 negligible compared to 1. Therefore the loop inductance is too small to account for the observed interference pattern. The device is thus dominated by the kinetic inductance of the weak-links. This is a common feature of SQUIDs based on microbridges of size much larger than the superconducting film coherence length ?[23]?.

Histograms of $I_s$ exhibits a distribution shape typically found in Josephson junctions [25] from which we derive a residual noise of $1.5 \times 10^{-4}$ ?$_?$/Hz$^{½}$ at 40mK. This first series of samples has a roughly comparable behavior to those similar in design, based on thicker Nb Dayem bridges and fabricated by electron beam lithography (EBL) [26].

In order to reduce further the weak-link size, we have tried another design. Starting from a SQUID fabricated using the method described above, a 15nm-wide single oxide line partially oxidizing the niobium layer is drawn across both Dayem microbridges [see Fig.1(b) and sketch Fig. 3(b)]. Such a buried nanostructure belongs to a other family of weak links known as a "variable thickness bridges" [23]. The resulting field-modulation of $I_s$ obtained for these devices (Fig 3, bottom) follows a pattern which differs from previously presented Dayem SQUIDs. In every tested device based on this second type of weak links, the modulation depth ?$I_s$/$I_s$ is enhanced by roughly a factor of two with respect to previous devices Furthermore, a clear deviation from the linear dependence of $I_s$ (??)is observed near the current maxima (Fig. 3b) . As pointed out in [23], these two phenomena are related to the smaller size of the weak links which characteristic dimensions become of the order of the superconducting film coherence length ??(estimated to be around 10nm).

The effective cross-section of the microbridges for that geometry is not exactly known but is estimated to be around 50nm wide and 3nm thick by measuring the niobium oxide height. Depending on the film thickness and quality, the critical current density in our junctions varied from 0.3 to 8 MA/cm².

We have checked that these devices have sufficiently high critical magnetic field in the in-plane direction to allow local magnetic flux detection. Indeed EBL-made Dayem SQUIDs have already enabled powerful magnetometry techniques in nanomagnetism [27], leading to measurements of magnetization reversal in nanoscale magnetic particles. AFM-made devices should offer new features such as the fabrication at a chosen position allowing an optimized coupling to magnetic signals, and an increased intrinsic sensitivity. In the case of small magnetic clusters which are placed very close to the microbridge junctions, we also expect an improvement of one to two orders of magnitude due to the reduction of the microbridges size. It might allow us to detect the spin flips of about 100 magnetic moments.

We are indebted to G. Battuz and A. Hadj-Azzem for substrate preparation and to A. Bychkov for ultra-thin film characterization. V. Safarov, F. Salvan and K. Hasselbach are gratefully acknowledged for help and discussions.

# Figure captions

Fig. 1 : Atomic Force micrographs of 6.5nm-thick Niobium strip lines on which SQUIDs has been made by AFM-controlled anodisation. Brightest regions are the patterned niobium oxide protrusions. In Fig. 1a, the microbridges on each side of the loop consists of 50nm wide 400nm long lateral constrictions, reminiscent to the "Dayem" type bridges. In Fig. 1b the bridges have a variable thickness made by oxide lines locally reducing the Niobium layer. 0.3-nm-high atomic steps and their replica can be seen on the bare sapphire substrate (dark lateral areas) as well as on top of both raw and patterned surfaces thus showing the structural quality of the deposited and of the oxidized layers.

Fig. 2 : Voltage-current characteristics measured at 40mK of the Dayem SQUID shown in Fig. 1a. The switching current $I_s$ is defined as the maximum observed Josephson current. Hysteresis observed between ramp-up and ramp-down is due to Joule heating in the strip line having discrete widths. Top inset : Magnetic field dependence of $I_s$ measured at 0.5K. Bottom inset: temperature dependence of the $I_s$ normalized histograms taken at 10kHz.

Fig. 3 : Schematics diagrams describing the two kinds of microbridges (the grey volume corresponds to formed oxide) and respectively measured flux modulation of the switching current (T=40mK) for SQUIDs based on each geometry.
(a): For SQUIDs with 0.3µm-long "Dayem"-like bridges, a perfect saw-tooth modulation is obtained, since bridges have width and length larger than the superconducting coherence length ?.
(b): For SQUIDs with variable thickness bridges, the magnetic field dependence of Is shows a deviation from the linear behavior (dotted line) while modulation depth is increased with respect to curve (a). These features appear as signatures of Josephson junctions of dimensions shorter than ?.

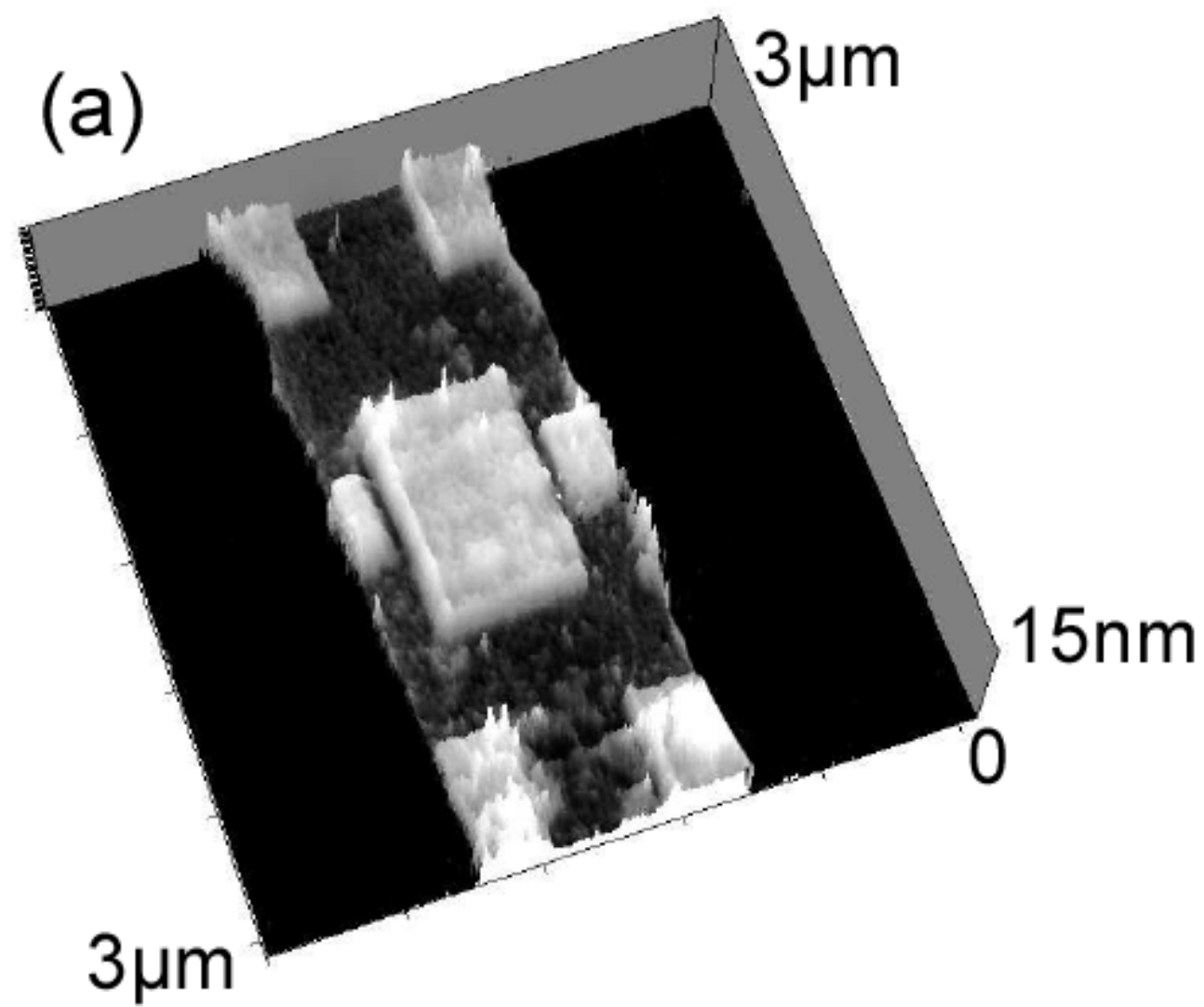

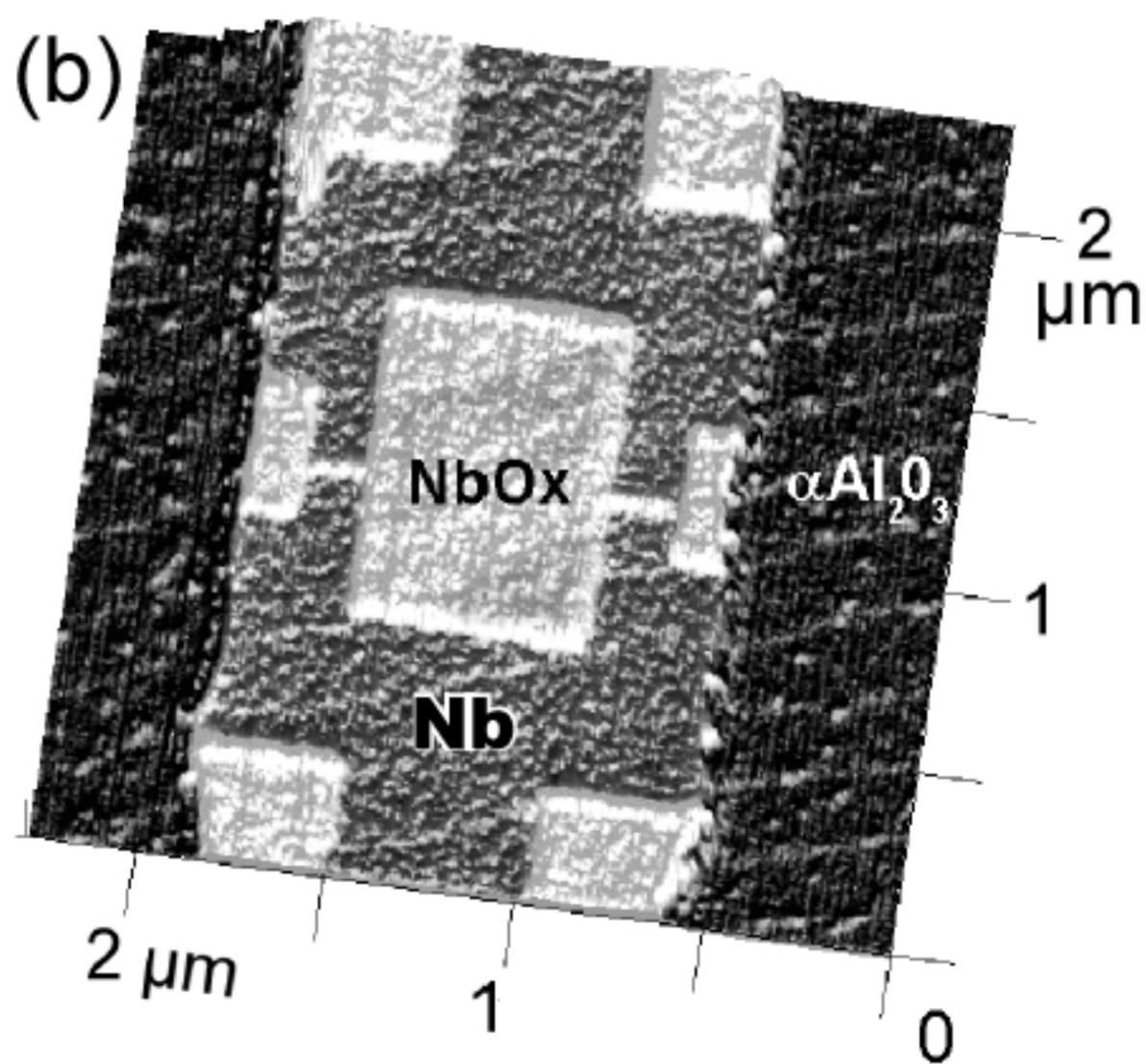

FIG. 1   Bouchiat et al.

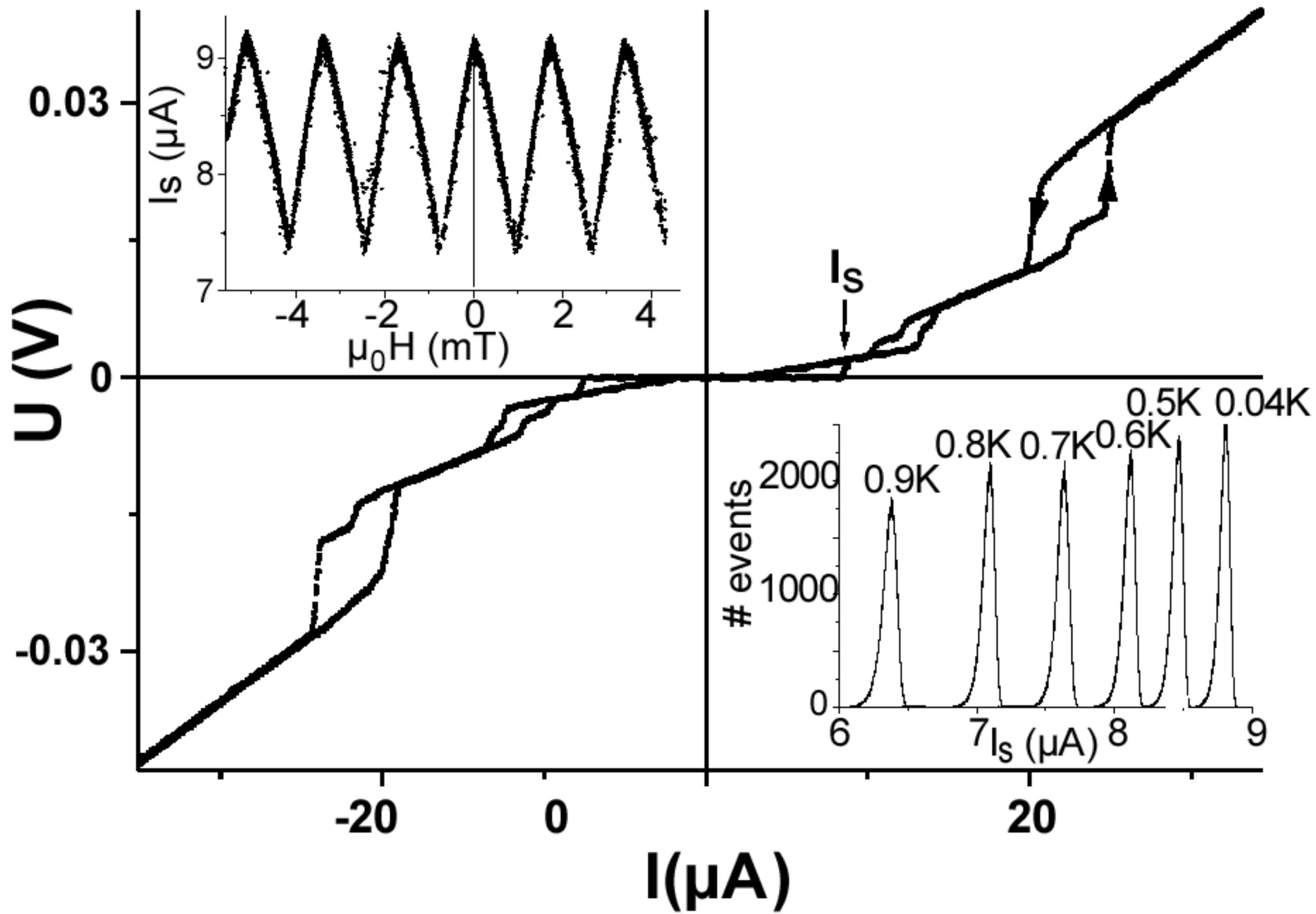

figure 2

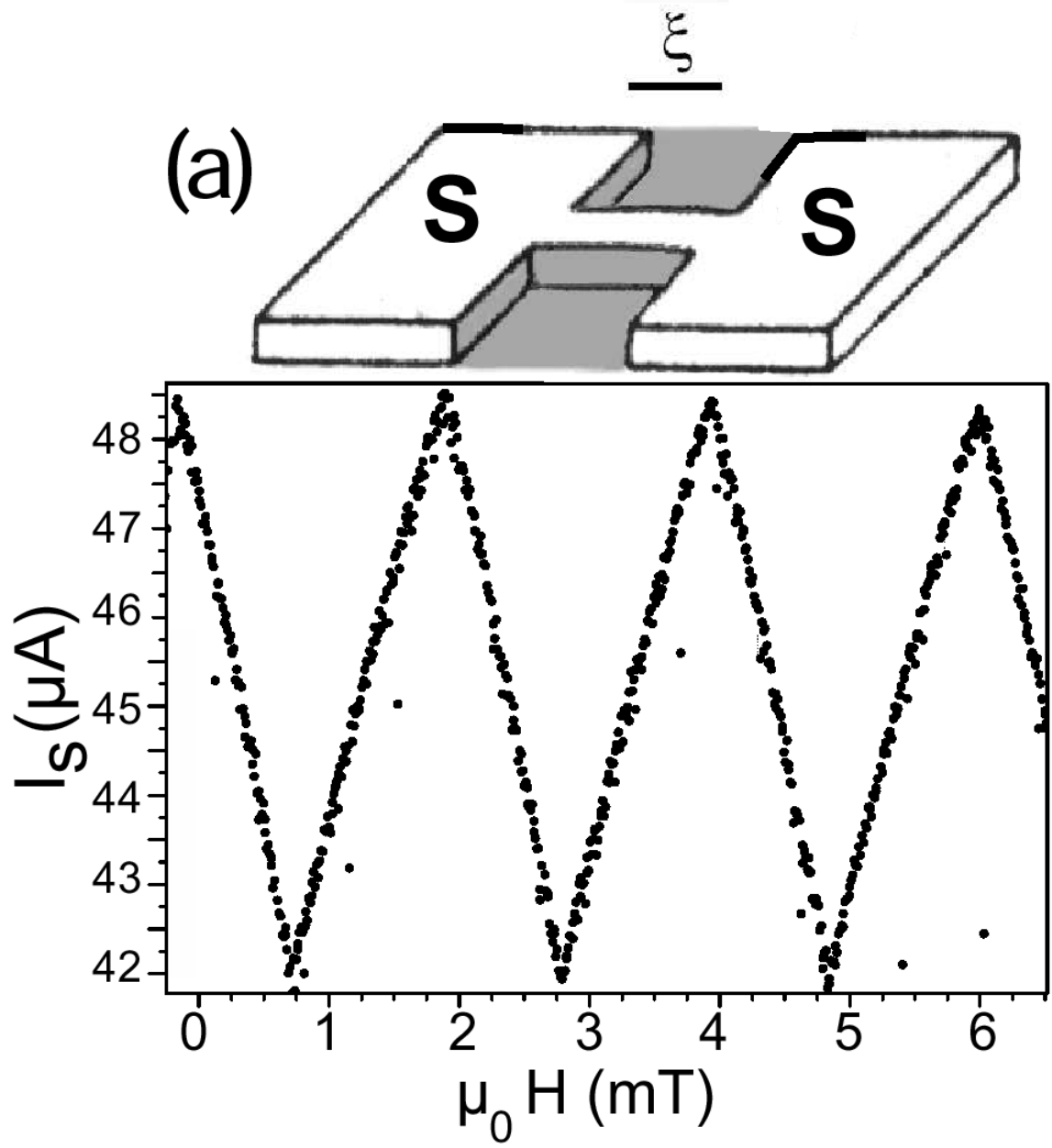
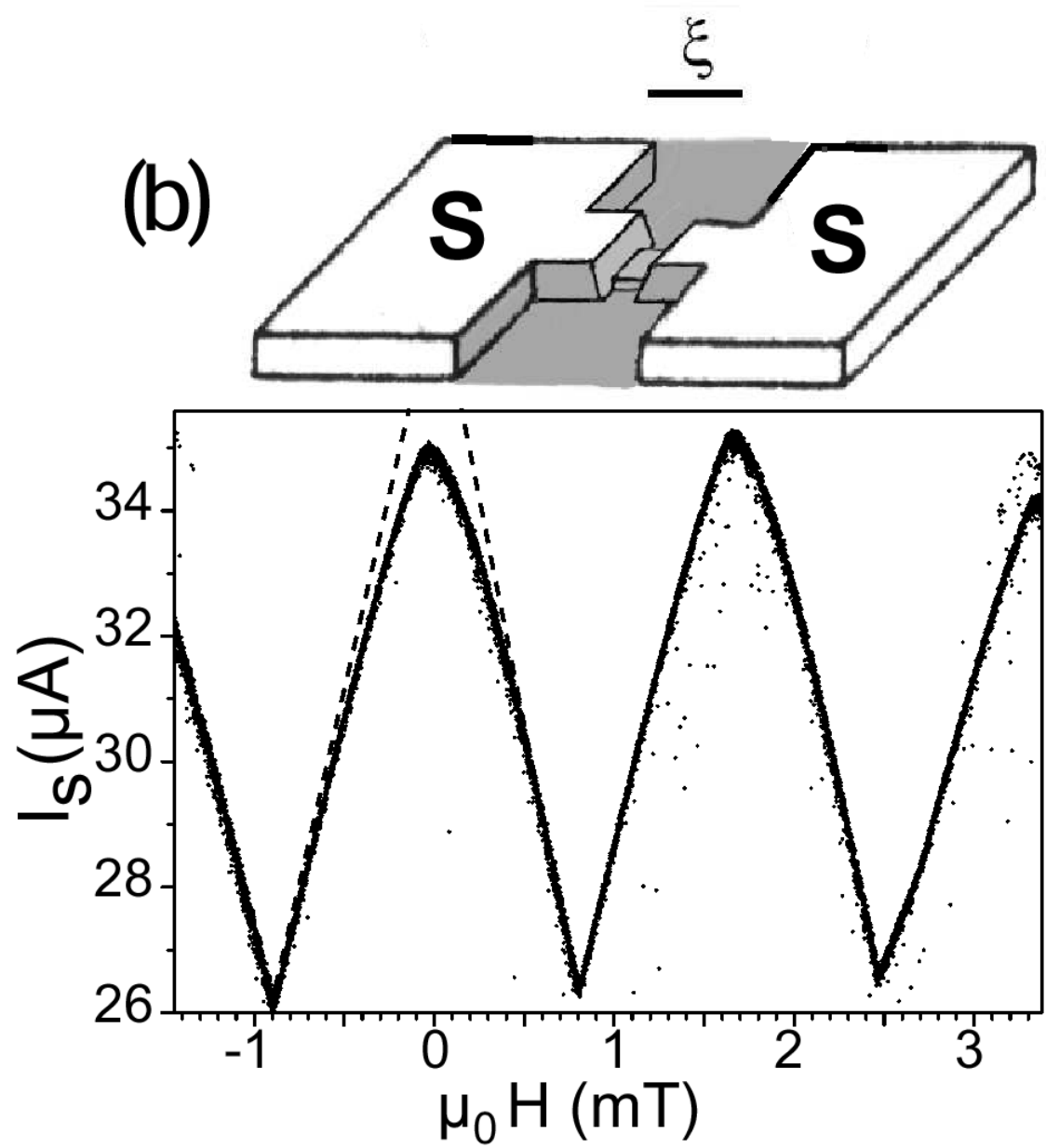

FIG. 3    BOUCHIAT et al.